\documentclass[
aps,%
11pt,%
final,%
notitlepage,%
oneside,%
onecolumn,%
nobibnotes,%
nofootinbib,%
superscriptaddress,%
noshowpacs,%
centertags]%
{revtex4}

\begin{document}
\selectlanguage{english}

\title{The Structure of Clusters with Bimodal Distributions of\\
Galaxy Radial Velocities. I. A1035}

\author{\firstname{A.~I.}~\surname{Kopylov}}
\affiliation{\saoname}

\author{\firstname{F.~G.}~\surname{Kopylova}}
\affiliation{\saoname}

\received{July 9, 2007}%
\revised{July 19, 2007}%

\begin{abstract}
The structure of the A1035 cluster of galaxies ($10^h32^m
+40^{0}13'$, $cz\sim 22000$~km\,s$^{-1}$), which exhibits a bimodal
distribution of galaxy radial velocities ($\Delta V\approx
3000$~km\,s$^{-1}$), is analyzed using three methods of determining
the relative distances to clusters from early-type galaxies: the
Kormendy relation corrected for the dependence of residuals on
galaxy magnitude, the photometric plane, and the fundamental
plane. We use the data obtained with the 1-m telescope of the
Special Astrophysical Observatory of the Russian Academy of
Sciences  and  SDSS (DR5) data to show that A1035 consists of two
gravitationally unbound independent clusters. These clusters with
the velocity dispersions of  566~km\,s$^{-1}$ and 610~km\,s$^{-1}$
and masses within $R_{200}$ equal to $2.7\cdot10^{14}$ and
$3.5\cdot10^{14}$ $M_{\odot}$, respectively, obey the Hubble law.
\end{abstract}
\maketitle

\section{INTRODUCTION}
Different methods of determination of masses of clusters of
galaxies  (based on x-ray flux, gravitational lensing of galaxies
by the cluster, or on the virial theorem) yield the results that
agree well with each other for the central regions of regular
clusters. The mass  distribution on scale lengths exceeding the
size of virialized cluster regions  (1--2~Mpc) is much more
poorly known. In this connection it is of interest to study the
dynamics of interaction of neighboring clusters and subclusters
within the same cluster. The velocity dispersion amounts to
$1000-1500$~km\,s$^{-1}$ in the richest clusters, where galaxies
sometime form complex structures
--- subclusters. Of special interest are the cases where the
velocity distribution in a cluster has a bimodal form. A
$3000-3500$~km\,s$^{-1}$ difference between the mean radial
velocities may be due to either the gravitational interaction
between extremely massive clusters colliding close to the line of
sight (see Hayashi and White\cite{Hayashi} for
theoretical estimates of limiting velocities in terms of the
$\Lambda CDM$), or to the line-of-sight projection of unbound
clusters.

Sufficiently reliable direct estimates of the distances to
subclusters have been obtained for two clusters with a bimodal
galaxy velocity distribution: A3526 (Centaurus)
\cite{Lucey} and A2626 \cite{Mohr}. In both
cases the subclusters are located at the same heliocentric
distance and the velocity difference (1500 and 2600~km\,s$^{-1}$,
respectively) is due to the gravitational interaction between the
subclusters. In the case of A2626 the higher subcluster velocity
difference compared to velocities of internal motions  may be due
to the large mass of the (dark) matter at the periphery of the
clusters \cite{Mohr}.

We selected a total of four rich clusters (A1035, A1569, A1775,
and A1831) with a bimodal distributions of galaxy radial velocities
($\Delta V\sim 3000$~km\,s$^{-1}$) for direct (i.e., redshift
independent)  determination of subcluster distances and for the
determination of the nature of interaction between the
subclusters. In this paper we determine the (line-of-sight)
structure  of the A1035 cluster using three different methods to
estimate relative distances from early-type galaxies. This work
makes use of observational material obtained with the 1-m
telescope of the Special Astrophysical Observatory of the Russian
Academy of Sciences and the data of the  SDSS (Sloan Digital Sky
Survey) catalog.

\begin{table*}[tbp]
\setcaptionmargin{0mm} \onelinecaptionstrue \captionstyle{normal}
\caption{Parameters of early-type galaxies based on the results
of observations made with the 1-m telescope}
\label{data:Kopylov_n}
\bigskip
\begin{tabular}{l|c|c|c|c|c|c|c|c} \hline
 Clluster& Galaxy& $\alpha~~(J2000)~~\delta$& $z_h$& $cz_h$,& $m_R$,& $R_e$,&     $\mu_e$,& $n$\\
       &No.&                          &      & km\,s$^{-1}$& mag&  arcsec& mag$/\sq''$&    \\
\hline
A1035A& 1& 10 32 23.47 +40 10 08.6& 0.067020 &20092 & 14.15& 10.33& 22.61& $1.92\pm0.46$\\
A1035A& 2& 10 32 15.27 +40 10 12.5& 0.067040 &20098 & 14.30& 14.06& 23.38& $4.95\pm1.26$\\
A1035A& 3& 10 32 28.89 +40 08 54.6& 0.068691 &20593 & 14.82&  4.55& 21.26& $1.29\pm0.19$\\
A1035A& 4& 10 31 55.98 +40 06 43.7& 0.068831 &20635 & 14.99&  3.09& 20.45& $2.52\pm1.64$\\
A1035A& 5& 10 32 05.95 +40 17 25.7& 0.072430 &21714 & 15.20&  3.84& 21.36& $2.07\pm0.24$\\
A1035A& 6& 10 32 12.49 +40 08 08.4& 0.065940 &19768 & 15.55&  3.60& 21.48& $1.52\pm0.15$\\
A1035A& 7& 10 32 19.90 +40 08 26.5& 0.065582 &19661 & 15.50&  3.27& 21.25& $2.48\pm0.54$\\
A1035A& 8& 10 31 37.06 +40 07 39.6& 0.071793 &21523 & 15.70&  2.94& 21.14& $1.88\pm1.93$\\
A1035A& 9& 10 31 22.79 +40 10 09.9& 0.068496 &20535 & 16.08&  2.24& 20.89& $1.15\pm0.11$\\
\hline
A1035B& 1& 10 32 13.92 +40 16 16.4& 0.077816 &23329 & 13.77& 11.64& 22.51& $4.91\pm0.71$\\
A1035B& 2& 10 31 57.03 +40 18 20.7& 0.078951 &23669 & 15.22&  3.96& 21.59& $3.67\pm0.43$\\
A1035B& 3& 10 32 07.42 +40 10 30.3& 0.076860 &23042 & 15.42&  4.06& 21.54& $2.48\pm0.51$\\
A1035B& 4& 10 32 10.77 +40 17 02.4& 0.081590 &24460 & 16.02&  2.77& 21.34& $2.50\pm1.51$\\
A1035B& 5& 10 31 37.75 +40 10 32.6& 0.077238 &23155 & 16.17&  2.72& 21.49& $1.34\pm0.62$\\
A1035B& 6& 10 32 07.52 +40 15 49.7& 0.074700 &22394 & 16.44&  2.21& 20.99&  ---         \\
A1035B& 7& 10 32 18.29 +40 17 01.5& 0.078633 &23574 & 16.43&  3.16& 21.90& $0.91\pm0.45$\\
\hline
\end{tabular}
\end{table*}

This paper has the following layout. Section I gives an
introduction. Section II describes samples of early-type
galaxies. Section III describes the specifics of the use of
samples for the determination of distances to clusters of
galaxies and their analysis. In conclusion, we list the main
results of this study. We adopt the following values for
cosmological parameters: $\Omega_m=0.3$, $\Omega_{\Lambda}=0.7$,
and $H_0=70$~km~s$^{-1}$~Mpc$^{-1}$.
\begin{figure*}[tbp]
\setcaptionmargin{5mm} \onelinecaptionsfalse
\includegraphics[scale=0.5,angle=-90]{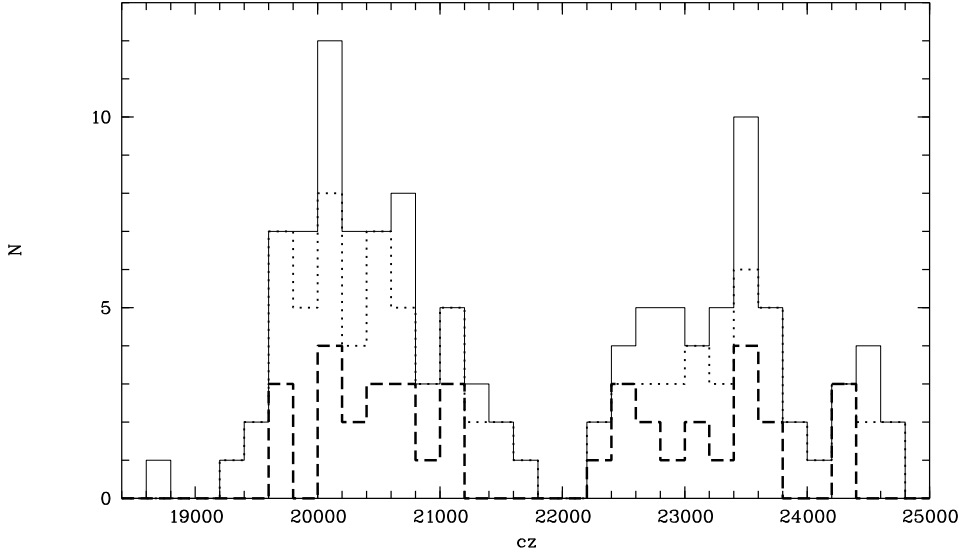} 
\captionstyle{normal} \caption{ Distribution of galaxy radial
velocities in the A1035 cluster inside the $30'$ (the solid line)
and inside the $R_{200}$ (the dotted line) clustercentric radius.
The dashed line shows the distribution of early-type galaxies
inside $R_{200}$.} \label{cz:Kopylov_n}
\end{figure*}

\begin{figure*}
\setcaptionmargin{5mm} \onelinecaptionsfalse
\includegraphics[scale=0.49,angle=-90]{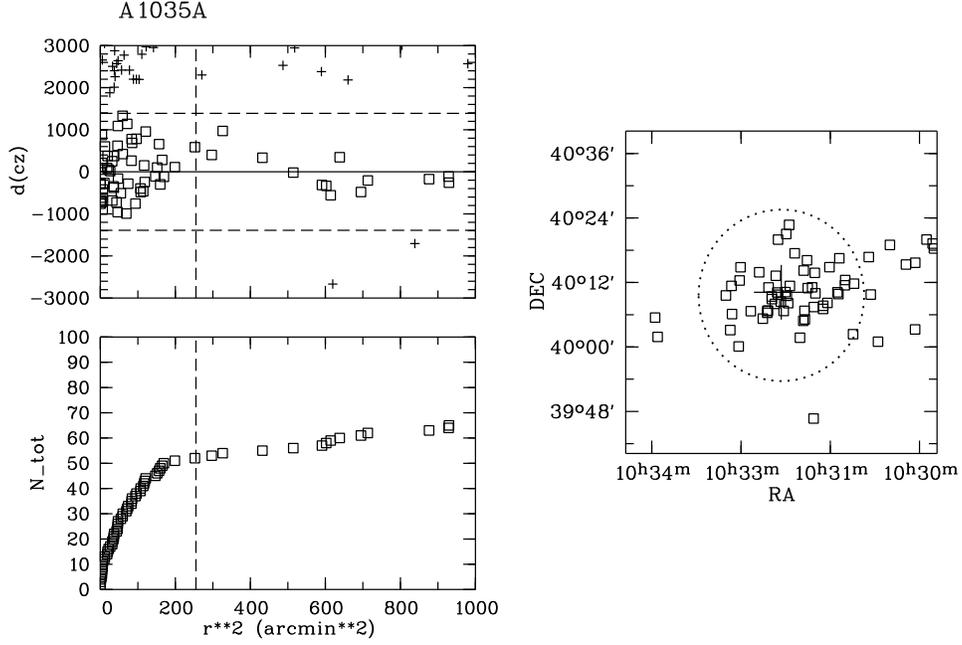}
\captionstyle{normal} \caption{ Distribution of galaxies in
A1035A. The top left panel shows the deviation of galaxy radial
velocities from the mean radial velocity: the dashed lines
correspond to the $\pm2.5\sigma$ deviation and the vertical
dashed line corresponds to the $R_{200}$ radius. The squares and
plus signs indicate cluster members and field galaxies,
respectively. The bottom left panel shows the integrated
distribution of the number of galaxies as a function of squared
clustercentric angular distance. The same designations are used.
The right-hand panel shows the sky-plane distribution of cluster
galaxies. The circle shows the  $R_{200}$-radius region in
arcmin. The cross indicates the cluster center.}
\label{paspA:Kopylov_n}
\end{figure*}

\begin{figure*}
\setcaptionmargin{5mm} \onelinecaptionstrue
\includegraphics[scale=0.5,angle=-90]{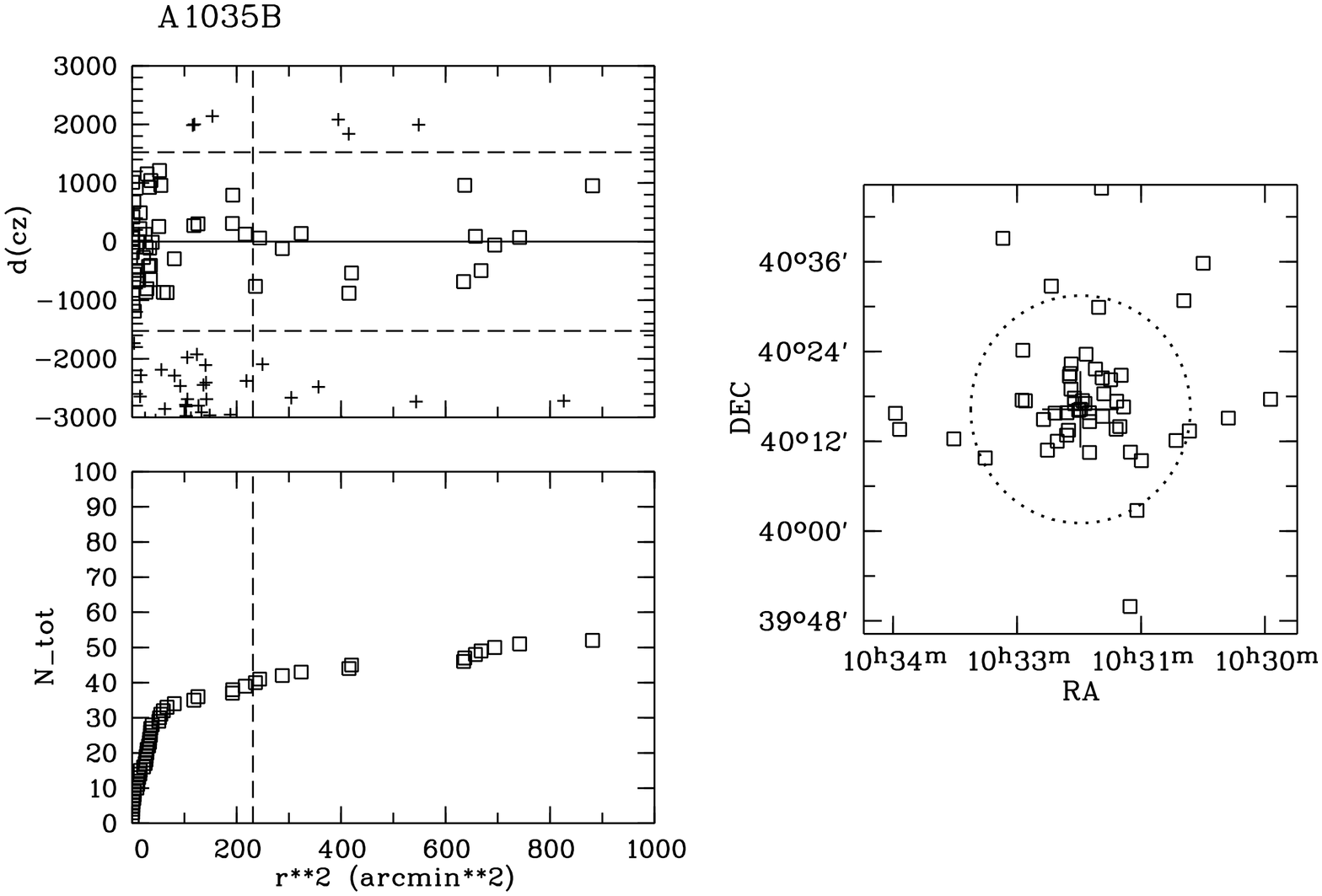} 
\captionstyle{normal}
\caption{Same as Fig.~2, but for A1035B.}
\label{paspB:Kopylov_n}
\end{figure*}

\section{DESCRIPTION OF DATA}
In this section we describe observational data
--- the parameters of early-type galaxies that we use to determine the relative
distances to the subsystems in the A1035 cluster
\cite{Abell}. According to Abell's catalog, this
cluster has a radial velocity of $cz\simeq 24000$ km/s, richness
class 2, and belongs to  Bautz-Morgan II--III type. We use SDSS
(DR5) data to identify two subsystems --- A1035A and A1035B ---
in the A1035 cluster (Fig.~\ref{cz:Kopylov_n} shows the velocity
distribution).Figures~\ref{paspA:Kopylov_n}
and~\ref{paspB:Kopylov_n} show our results for the main
parameters of the clusters: deviations of the radial velocities
of cluster members from the mean radial velocity of the cluster;
integrated distribution of clustercentric distances of cluster
members, and the sky-plane locations of cluster members. The
center of A1035B coincides with the position of the brightest cD
galaxy and with the peak of x-ray flux distribution. We adopt the
center of A1035A to be located between the two brightest galaxies.

\setcaptionwidth{\linewidth}%
\setcaptionmargin{0mm}%
\onelinecaptionstrue%
\captionstyle{normal}%
\begin{longtable*}{c|c|c|c|c|c|c|c|c|c}
\caption{Parameters of early-type SDSS galaxies}
\label{sdss:Kopylov_n}\\
 \hline
 Cluster.& $\alpha~~(J2000)~~\delta$& $z_h$& $cz_h$,&   $\sigma$,&   $r$,& $R_e$,&
$fracDeV_r$& $r90/r50$& $eClass$ \\
       &   &    &km\,s$^{-1}$& km\,s$^{-1}$& mag&  arcsec&  &   &  \\
\hline
\endfirsthead
\caption{
 (Contd.)}\\  
\hline Cluster& $\alpha~~(J2000)~~\delta$& $z_h$& $cz_h$,&
$\sigma$,& $r$,& $R_e$,&
$fracDeV_r$& $r90/r50$& $eClass$ \\
       &    &     &km\,s$^{-1}$& km\,s$^{-1}$& mag& arcsec& &     &       \\
\hline
\endhead
\hline
\endfoot
\hline 
\endlastfoot
A1035A & 10 32 15.27 +40 10 12.5& 0.067102& 20117& 208& 15.18& 4.24& 1.000& 3.49& -0.159\\
A1035A & 10 32 28.89 +40 08 54.6& 0.066943& 20069& 196& 15.22& 5.34& 0.907& 2.93& -0.142\\
A1035A & 10 32 23.47 +40 10 08.6& 0.066900& 20056& 220& 15.26& 4.47& 1.000& 3.38& -0.162\\
A1035A & 10 31 55.98 +40 06 43.7& 0.069003& 20687& 240& 15.30& 3.05& 0.980& 3.11& -0.143\\
A1035A & 10 32 33.51 +40 06 41.6& 0.069241& 20758& 199& 15.52& 3.65& 1.000& 3.28& -0.132\\
A1035A & 10 32 12.49 +40 08 08.4& 0.065940& 19768& 177& 15.80& 4.24& 0.952& 3.18& -0.137\\
A1035A & 10 32 19.90 +40 08 26.5& 0.065582& 19661& 208& 15.85& 2.80& 1.000& 3.17& -0.142\\
A1035A & 10 33 00.49 +40 14 50.3& 0.070582& 21160& 163& 16.14& 2.85& 1.000& 3.26& -0.122\\
A1035A & 10 31 52.75 +40 10 56.7& 0.068040& 20398& 190& 16.18& 2.78& 0.918& 2.90& -0.156\\
A1035A & 10 32 24.94 +40 13 14.6& 0.067106& 20118& 148& 16.33& 3.06& 0.904& 3.02& -0.140\\
A1035A & 10 32 22.64 +40 19 58.5& 0.070606& 21167& 129& 16.36& 2.82& 1.000& 2.62& -0.152\\
A1035A & 10 31 22.79 +40 10 09.9& 0.068496& 20535& 164& 16.42& 1.50& 1.000& 2.99& -0.136\\
A1035A & 10 31 56.77 +40 14 12.7& 0.067393& 20204& 124& 16.46& 2.74& 0.859& 2.88& -0.129\\
A1035A & 10 31 14.93 +40 12 26.9& 0.070187& 21042& 125& 16.54& 2.15& 1.000& 3.12& -0.126\\
A1035A & 10 32 17.02 +40 06 42.1& 0.070001& 20986& 132& 16.95& 2.17& 1.000& 2.91& -0.128\\
A1035A & 10 31 20.88 +40 16 30.6& 0.068932& 20665& 115& 16.96& 1.73& 0.906& 2.73& -0.123\\
A1035A & 10 31 45.59 +40 13 47.7& 0.065967& 19776& 147& 17.07& 1.74& 0.908& 2.84& -0.150\\
A1035A & 10 32 33.59 +40 06 21.4& 0.068162& 20434& 114& 17.10& 2.72& 0.831& 2.83& -0.125\\
A1035A & 10 31 06.04 +40 11 46.5& 0.068363& 20495& 151& 17.15& 1.36& 0.978& 2.72& -0.132\\
\hline
A1035B &10 31 57.03 +40 18 20.7& 0.079322& 23780& 218& 15.56& 4.25& 1.000& 3.31& -0.143\\
A1035B &10 31 04.63 +40 12 06.9& 0.080857& 24240& 146& 15.84& 4.89& 1.000& 2.95& -0.140\\
A1035B &10 32 54.02 +40 17 24.3& 0.075323& 22581& 219& 16.31& 1.85& 1.000& 3.34& -0.134\\
A1035B &10 32 10.77 +40 17 02.4& 0.081375& 24396& 153& 16.33& 3.20& 0.848& 2.93& -0.128\\
A1035B &10 31 47.68 +40 17 21.7& 0.075329& 22583& 170& 16.36& 3.17& 1.000& 2.94& -0.157\\
A1035B &10 31 48.12 +40 13 36.7& 0.076802& 23025& 134& 16.45& 2.00& 1.000& 3.01& -0.138\\
A1035B &10 32 18.29 +40 17 01.5& 0.078633& 23574& 121& 16.71& 3.36& 0.997& 2.64& -0.137\\
A1035B &10 32 07.52 +40 15 49.7& 0.074700& 22394& 195& 16.74& 2.12& 0.817& 2.68& -0.136\\
A1035B &10 32 22.67 +40 13 28.5& 0.078169& 23434& 159& 16.92& 1.39& 1.000& 2.98& -0.126\\
A1035B &10 33 23.12 +40 09 46.5& 0.078639& 23575& 141& 16.97& 1.93& 1.000& 2.83& -0.112\\
A1035B &10 32 56.32 +40 17 29.7& 0.075317& 22579& 153& 17.02& 1.21& 1.000& 3.04& -0.124\\
A1035B &10 31 52.13 +40 20 13.4& 0.081302& 24374& 122& 17.10& 3.44& 0.974& 2.82& -0.106\\
A1035B &10 32 20.82 +40 18 58.5& 0.076050& 22799& 122& 17.17& 2.00& 0.831& 2.67& -0.137\\
A1035B &10 31 58.21 +40 20 28.3& 0.077899& 23354& 145& 17.44& 1.40& 0.947& 2.69& -0.136\\
A1035B &10 32 22.02 +40 20 40.8& 0.077310& 23177& 114& 17.45& 1.41& 0.964& 2.66& -0.097\\
A1035B &10 32 38.02 +40 10 48.7& 0.079071& 23705& 139& 17.45& 1.74& 1.000& 2.72& -0.099\\
A1035B &10 32 21.50 +40 21 02.7& 0.078636& 23574& 102& 17.47& 1.52& 1.000& 2.93& -0.021\\
A1035B &10 32 23.81 +40 15 49.0& 0.076353& 22890& 151& 17.47& 1.54& 1.000& 3.03& -0.111\\
A1035B &10 32 40.81 +40 14 55.0& 0.075532& 22644& 117& 17.69& 1.28& 0.909& 2.77& -0.104\\
\end{longtable*}

\begin{figure*}
\setcaptionmargin{5mm} \onelinecaptionsfalse
\includegraphics[scale=0.6,angle=-90]{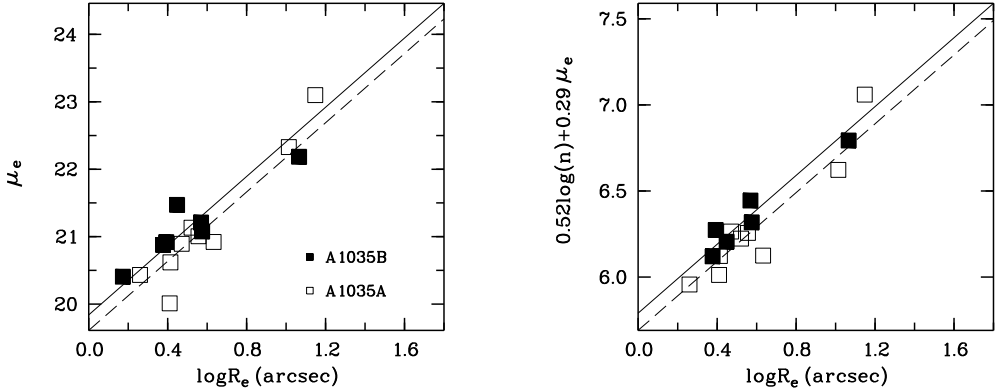} 
\captionstyle{normal}
\caption{ The Kormendy relation (the
left-hand panel) and photometric plane (the right-hand panel) for
early-type galaxies in  A1035A and A1035B based on the results of
observations made with the 1-m telescope. The dashed and solid
lines indicate the zero points of clusters A and B, respectively.}
\label{kr:Kopylov_n}
\end{figure*}

Table~\ref{cluster:Kopylov_n} lists the cluster parameters that
we determined for the objects located within the $R_{200}$ radius
according to SDSS data. Here $R_{200}$ is the radius of the
virialized part of the cluster where the mass density exceeds the
critical density of the Universe by a factor of 200. In this
case, the cluster mass can
be determined from $R_{200}$ and velocity dispersion $\sigma$.
We determined the average radial velocity  $cz$ of the cluster
and its dispersion $\sigma$ iteratively: we first use all
galaxies with measured radial velocities ($N=65$ and $N = 52$ in
A1035A and A1035B, respectively) inside the region of radius
$30'$ studied here except for those that deviate by more than
2.5~$\sigma$. We further assume that the clusters are in virial
equilibrium and that their masses increase linearly with radius
to compute, with the resulting velocity dispersion, the radius
$R_{200}$~ = $\sqrt{3}\sigma(1+z)^{-3/2}/(10H_0)$\,Mpc
\cite{Carlberg} and redetermine the average cluster
radial velocity  $cz$ and its dispersion $\sigma_{200}$ inside the
$R_{200}$ radius. The virial mass within  this radius is equal to
$M_{{vir},{200}} = 3G^{-1}R_{200}\sigma_{200}^{2}$. Here
$N_{200}$ is the number of galaxies with measured radial
velocities located inside $R_{200}$. We adopt the x-ray
luminosities from Rines and Diaferio~\cite{Rines}. The
same authors showed that A1035 consists of two subclusters.
Figure~\ref{cz:Kopylov_n} shows the distribution of galaxy radial
velocities in the systems inside the selected radius.

\begin{table*}[tbp]
\setcaptionmargin{0mm} \onelinecaptionstrue
\captionstyle{flushleft} \caption{Cluster data}
\label{cluster:Kopylov_n}
\medskip
\begin{tabular}{l|c|c|c|c|c|c|c|c} \hline
 Cluster& $\alpha~~(J2000)~~\delta$& $z_h$&$cz_h$,& $   \sigma$,& $R_{200}$,& $N_{200}$
& $M_{200}$,          & $L_x$,               \\
       &                          &   & km\,s$^{-1}$   & km\,s$^{-1}$& Mpc      &
& $10^{14}~M_{\odot}$& $10^{43}$~erg\,s$^{-1}$\\
\hline
 A1035A & 10 32 19.36 +40 10 10.4 &0.067992&20383  & $556\pm77$ &1.24 &52&$2.68\pm0.74$& 0.7\\
 A1035B & 10 32 13.95 +40 16 16.5 &0.078216&23448  & $610\pm98$ &1.35 &39&$3.52\pm1.13$& 2.0\\
\hline
\end{tabular}
\end{table*}

\subsection{The Data for Early-Type Galaxies Based on Observations Made with the
1-m Telescope} The relative distances of clusters of galaxies in
sufficiently distant regions with $z=0.05$ and higher are
determined using parameters of early-type galaxies (e.g.,
\cite{Colless}; \cite{Hudson}). In this paper
we use three methods based on the properties of early-type
galaxies to estimate the peculiar motions of clusters A1035A and
A1035B: the Kormendy relation \cite{Kormendy}, the
photometric plane (PhP) \cite{Graham}, and the
fundamental plane (FP) \cite{Djorgovski}. To accomplish
our task, we determined the parameters of 16 galaxies in the
systems studied using the direct  $R_c$-band (the Kron--Cousins
system) images that we took with the 1-m telescope of the Special
Astrophysical Observatory of the Russian Academy of Sciences in
1998, 1999, and 2003 under average seeing conditions
($1.5\arcsec$) measured as the  FWHM of a star's profile. In 1998
and 1999 a $520\times580$  ISD015A CCD was used with a pixel size
of $18\times24$ $\mu$m, which corresponds to an angular size of
$0.28\arcsec\times 0.37\arcsec$. In 2003, a $2048\times2048$ CCD
with an angular pixel size of  $0.43\arcsec$ was used. The
exposures were equal to 500 or 600 s. Landolt's
\cite{Landolt} standard stars were observed several
times during each night to provide photometric calibration.

We use MIDAS (Munich Image Data Analysis System,
\cite{Grosbol}) to reduce the observational data. We
apply the standard procedure of image reduction: subtraction of
median dark frame, division by flat field, and subtraction of the
sky background approximated by a quadric surface. We use
multiaperture photometry to determine the total asymptotic
magnitude of each galaxy. We then use the total magnitude to
determine the effective radius $R_e$ of the circle where the
galaxy luminosity decreases by a factor of two, and the effective
surface brightness $\mu_e$ at this radius. We determine the
exponent $n$ characterizing the shape of the surface brightness
profile by fitting a Sersic \cite{Sersic} profile
$R^{1/n}$ ($n$ = 4 for the de Vaucouleurs \cite{deV}
profile) to the observed profile in the galactocentric radius
interval from 3~FWHM out to the radius where the surface brightness
equals $24^m-25^m$~arcsec$^{-2}$. We then use the method proposed
by Saglia et al. \cite{Saglia} to correct the resulting
photometric parameters of galaxies  ($R_e$ and $\mu_e$), except
for $n$. We compare independent measurements for 15 galaxies that
we observed twice to find that the standard error of measured
$\mu_e$ and $\log(R_e)$ is equal to $0.^m09$ and 0.02,
respectively. We thus use model-independent galaxy
parameters($R_e$, $\mu_e$) estimated from the total asymptotic
absolute magnitude, and
model-dependent quantity $n$. The asymptotic magnitude was
difficult to determine for the three brightest galaxies (of very
large size), and that is why we estimate all parameters by fitting
Sersic's profile to observations.

Table~\ref{data:Kopylov_n} features the results of our photometric
measurements. It contains the following data (the observed galaxy
parameters are not corrected for seeing): the number of the
cluster according to Abell's catalog; the number of the galaxy;
the J2000 equatorial coordinates of galaxies; the heliocentric
redshift and radial velocity (according to NED); the total
(asymptotic) magnitude; the effective radius of the galaxy in
arcseconds; the effective surface brightness at the effective
radius, and the Sersic profile shape parameter $n$ and its error.

\begin{figure}[]
\centering
\setcaptionmargin{5mm}
\onelinecaptionsfalse
\includegraphics[scale=0.75,angle=-90]{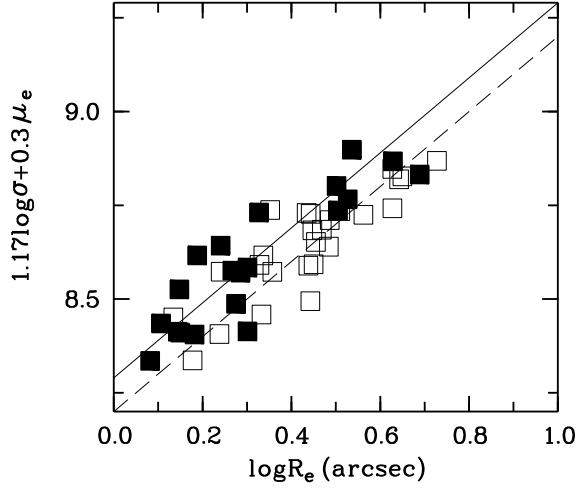}
\captionstyle{normal}
 \caption{ Fundamental plane for early-type galaxies in
A1035A and A1035B (SDSS data). Designations are the same as in Fig.~4.}
\label{fp:Kopylov_n}
\end{figure}

\begin{figure}[]
\setcaptionmargin{8mm}
\onelinecaptionsfalse
\includegraphics[scale=0.65,angle=-90]{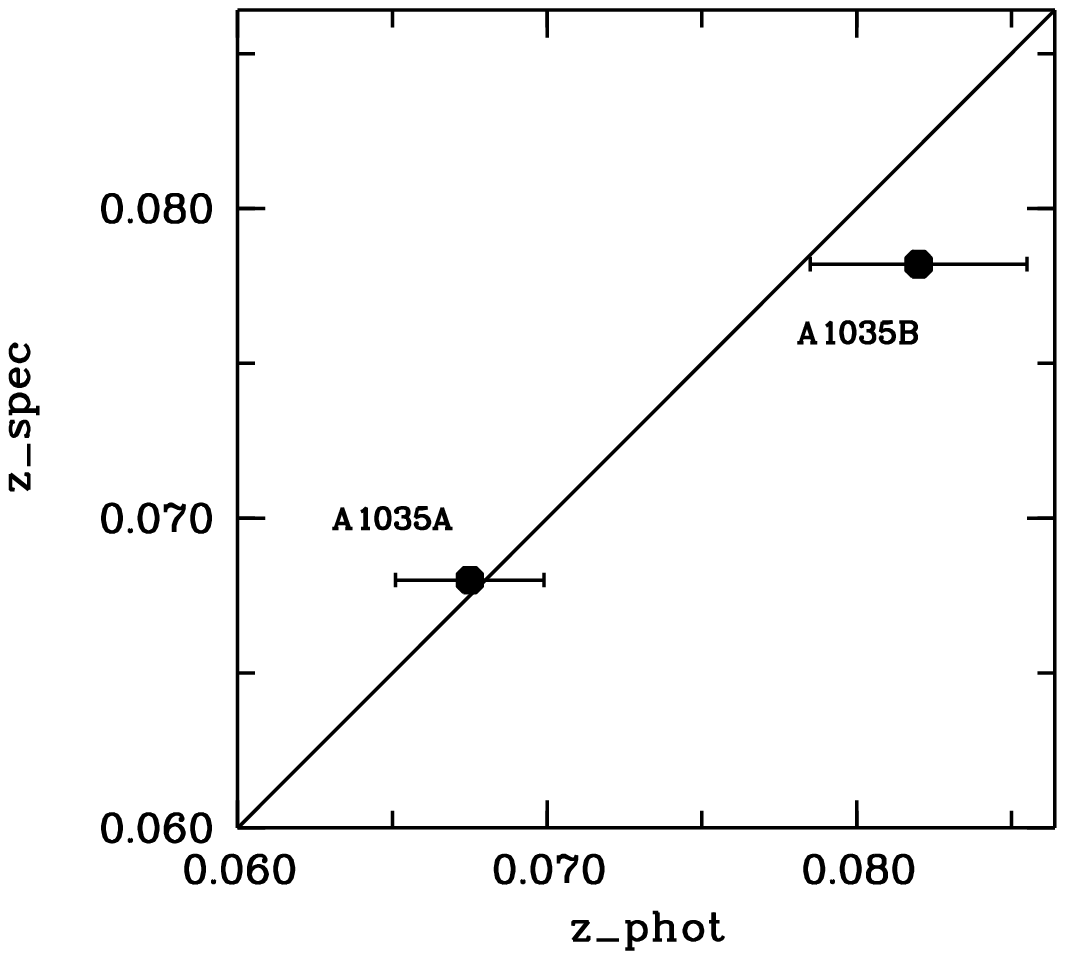}
\captionstyle{normal}
 \caption{The Hubble diagram for A1035A and
1035B clusters. The quoted errors correspond to the error of the
average distance to the cluster.} \label{Hdiag:Kopylov_n}
\end{figure}

\subsection{Data for Early-type Galaxies from the SDSS Catalog}

We compiled a sample of early-type galaxies in the A1035A and
A1035B clusters based on  SDSS DR5 \cite{Adelman} data
($r$-band). We selected galaxies based on the criteria proposed by
Bernardi et al. \cite{Bern1} (down to the SDSS limiting
magnitude of  $17^m.77$). We found a total of 19 galaxies within
the virialized region in each cluster. Table \ref{sdss:Kopylov_n}
gives the following parameters for the sample of early-type
galaxies that we selected from the catalog: the J2000 equatorial
coordinates; the heliocentric redshift and radial velocity; the
central stellar velocity dispersion $\sigma$; the parameters of
the de Vaucouleurs profile (total magnitude and effective radius);
$fracDeV_r\geq 0.8$, the quantity that characterizes the
contribution of the de Vaucouleurs bulge to the surface brightness
profile of the galaxy; $r_{90}/r_{50}\geq 2.6$, the concentration
index, which is equal to the ratio of the radii containing 90\%
and 50\%  of the Petrosian flux, and $eClass\leq 0$, the parameter that
characterizes the spectrum of the galaxy: minus means that the
spectrum exhibits no appreciable emission lines.

\section{ANALYSIS OF RELATIVE DISTANCES OF THE SUBSYSTEMS IN A1035}
Redshift-independent methods of distance determination for
clusters of galaxies usually combine distance-dependent ($R_e$)
and distance-independent ($\mu_e$, $\sigma$) parameters of
early-type galaxies. In the case of the A1035 cluster considered
here, which has a bimodal distribution of radial velocities, there are
two possible variants: either the A1035A and A1035B subclusters
are gravitationally bound, located at the same distance, and
constitute a single large cluster, or they are gravitationally
unbound and obey the Hubble law, which relates radial velocity and
distance, and these two subclusters are independent clusters.

A detailed description of the determination of cluster distances
from the Kormendy \cite{Kormendy} relation corrected
for the dependence of residual velocities on galaxy magnitude can
be found in our earlier paper \cite{Kop1}. The relation
has the form: $\log R_e=0.38\mu_e+\gamma$.
Figure~\ref{kr:Kopylov_n} (left panel) shows this relation for
nine observed galaxies in A1035A and seven galaxies in A1035B. The
figure gives our estimates for the following parameters: the
seeing-corrected $\log R_e$ in arcsec and the surface brightness
values with cosmological correction $10\log(1+z)$. The zero point
of the relation varies with the distance to the galaxy and and is
assumed to be barely affected by other factors (e.g.,
metallicity). We find the following zero-point values with
magnitude correction applied: $\gamma_A=-7.4717~(rms=0.1641)$,
$N=9$; $\gamma_B=-7.4633~(rms=0.0474)$, and $N=7$. The zero-point
difference is equal to  $\gamma_{AB}=-0.008\pm0.058$, or, if
computed without the three brightest galaxies, to
$\gamma_{AB}=+0.058\pm0.038$. If the subsystems obey the Hubble
law, then the difference of radial velocities would imply a
zero-point difference of 0.061.

Photometric plane (PhP) can be derived from the fundamental plane
(FP) for early-type galaxies by substituting the photometrically
measurable Sersic-profile parameter for the spectroscopically
measurable parameter --- the central stellar velocity dispersion
in the galaxy. PhP was constructed, e.g., by Graham
\cite{Graham}. To construct it, we use the photometric
data ($R_e$, $\mu_e$) \cite{Kopyl} for 12 early-type
galaxies obtained with the 6-m telescope with 200-s exposures
under $1''$ seeing conditions. We determine Sersic's parameter
$n$ from the surface brightness profile.

If expressed in terms of $\log R_{e}$, the PhP has the following
form: $\log R_{e}= 0.521(\pm0.130)\log
n+0.291(\pm0.026)\mu_{e}+\gamma$. Figure\,\ref{kr:Kopylov_n} (right
panel) shows the photometric planes for the galaxies studied. We
find the following zero-points for the subsystems in A1035:
$\gamma_A$ = \mbox{--5.6905~~($rms$=0.1190)},~~ $N=9$; $\gamma_B$=
\mbox{--5.7878~~($rms$ = 0.0710)}, $N=7$. The zero-point
difference is equal to $\gamma_{AB}=+0.097\pm0.049$ or
$\gamma_{AB}=+0.129\pm0.048$ if computed without the three
brightest galaxies.

SDSS data for a greater number of galaxies allow the zero points
(cluster distances) to be estimated more accurately, because the
statistical accuracy depends on the number of galaxies. To
construct the FP, we compute the mean effective surface
brightness by the following formula: $<\mu_e>$= r+2.5 $\log(2 \pi
R_e^2)-10 \log(1+z)$. Central velocity dispersion $\sigma$ is
adjusted to the standard $1/8 R_e$ circular aperture in
accordance with \cite{Bern1}. Figure~\ref{fp:Kopylov_n}
shows the FP for the selected 38 galaxies with the coefficients
(direct regression in terms of  $\log R_e$) from Bernardi et al.
\cite{Bern2}, where it has the following form:
$\log R_{e}= 1.17\log \sigma+0.30 <\mu_{e}>+\gamma$. We find the
following zero points for the subsystems in A1035:
$\gamma_A=-8.2044~(rms=0.0683)$, $N=19$;
$\gamma_B=-8.2868~(rms=0.0844)$, and $N=19$. The zero-point
difference is equal to $\gamma_{AB}$=+0.082$\pm0.025$. The two
regressions (direct and orthogonal) yield an average zero-point
difference of $\gamma_{AB}=+0.076\pm0.026$, i.e., the distances to
the two clusters differ by almost $3\sigma$. Thus all
distance-measurement methods applied show that the subsystems in
the A1035 have not segregated from the Hubble flow and are
independent clusters. We also determine the peculiar velocities
of A1035A and A1035B in the same coordinate system as we used in
our earlier paper \cite{Kop2} with respect to the
common zero points equal to --8.093 and --8.807, respectively.
The less distant  A1035A and more distant A1035B clusters have
the peculiar velocities of $+148\pm730$~km\,s$^{-1}$ and
$-1112\pm1050$~km\,s$^{-1}$, respectively.
Figure\,\ref{Hdiag:Kopylov_n} shows the Hubble diagram (for the
most accurate method described above) for the two clusters
studied. We compute the photometric redshifts $z_{phot}$ (0.067532
and 0.081966 for A1035A and A1035B, respectively) corresponding
to our estimated cluster distance from the difference between the
common zero point and the zero point of each system.

\section{CONCLUSIONS}
We determine the $R_c$-band photometric parameters ($m_R$,
$\mu_e$, $\log(R_e)$, $n$) for 16 early-type galaxies in the
A1035 cluster with a bimodal distribution of radial velocities (the
cluster consists of two subclusters A1035A and A1035B) from the
observations made with the 1-m telescope of the Special
Astrophysical Observatory of the Russian Academy of Sciences. We
use these data to construct the Kormendy relation and the
$R_c$-band photometric plane for early-type galaxies. We use SDSS
(DR5) data to determine the main parameters of these clusters and
construct the $r$-band fundamental plane for early-type galaxies.
The distances to the clusters that we determine using the methods
described allow us to more accurately assess the dynamical state
of A1035 and determine the peculiar velocities of its subsystems.
Our main conclusion is that the A1035 cluster consists of two
independent systems located at their own Hubble distances. The
mass of the central virialized regions is insufficient to
gravitationally bind the clusters given the $\sim$3000 $km
s^{-1}$ difference of their radial velocities.

\begin{acknowledgments}

\vspace*{0.5cm} This work was supported in part by the Russian Foundation for
Basic Research (grant no.~07-02-01417a).

This research has made use the NASA/IPAC Extragalactic Database (NED),
which is operated by the Jet Propulsion Laboratory, California Institute
of Technology, under contract with the NASA.\\
Funding for the creation and distribution of the SDSS Archive has been
provided by the Alfred P. Sloan Foundation, the Participating Institutions,
the National Aeronautics and Space Administration, the National Science
Foundation, the US Department of Energy, the Japanese Monbukagakusho, and
the Max Planck Society. The SDSS Web site is http://www.sdss.org/.
\end{acknowledgments}

\begin{center}
\refname
\end{center}

\end{document}